\documentclass[twocolumn,showpacs,preprintnumbers,amsmath,amssymb]{revtex4}
\usepackage{graphicx}
\usepackage{dcolumn}
\usepackage{bm}
\usepackage{subfigure}
\begin{document}

\preprint{APS/123-QED}

\title{Gutzwiller study of extended Hubbard models with fixed boson densities
}

\author{Takashi Kimura}
\affiliation{Department of Information Sciences, 
Kanagawa University, 2946 Tsuchiya, Hiratsuka, Kanagawa 259-1293, Japan}
%

\date{\today}

\begin{abstract}
We studied all possible ground states, including supersolid (SS) phases 
and phase separations of hard-core- 
and soft-core-extended Bose--Hubbard models 
with fixed boson densities
by using the Gutzwiller variational wave function and the 
linear programming method. 
We found that the phase diagram of the soft-core model 
depends strongly on its transfer integral. 
Furthermore, for a large transfer integral, 
we showed that an SS phase 
can be the ground state even below or at half filling against 
the phase separation.  
We also found that the density difference 
between nearest-neighbor sites, 
which indicates the density order of the SS phase,  
depends strongly on the boson density and transfer integral. 
\end{abstract}

\maketitle

\section{Introduction}

The supersolid (SS) phase that 
simultaneously exhibits a superfluid (SF) phase 
and a density-wave order   
has been studied, since it was first 
proposed theoretically \cite{Andreev,Chester,Leggett}. 
Although its existence is still controversial,  
a non-classical moment of inertia of solid ${}^4$He 
in Vycor glass that suggests its existence
was recently reported \cite{Kim}. 

SS phases may also exist in optical lattices. 
Cold atoms in optical lattices are new experimental 
systems in which theoretically modeled Hamiltonians 
can be more distinctly simulated experimentally with no disorders. 
For instance, an optical lattice system, as described 
by a Bose--Hubbard model \cite{Fisher1,Jaksch}, distinctly showed  
the phase transition between the SF phase and 
the Mott insulator (MI) phase \cite{Greiner}. 
Moreover, recent observations of the 
long-range dipole--dipole interaction in 
${}^{52}$Cr atoms \cite{Cr} may lead to the realization of the SS phase. 

The simplest lattice model with a long-range interaction 
is the extended Bose--Hubbard model with a nearest-neighbor interaction: 
\begin{eqnarray}
H&\ = \ &H_{kin}+H_{int}^U+H_{int}^{V},\nonumber\\
H_{kin}&\ = \ &-t\sum_{\langle i,j \rangle}
(a_{i}^\dagger a_{j}^{}
 + a_{j}^\dagger a_{i}), \nonumber\\
H_{int}^U&\ = \ &\frac{U}{2}\sum_i n_i(n_i-1),\nonumber\\
H_{int}^V&\ = \ &V\sum_{\langle i,j \rangle} n_in_j. \label{Hamiltonian}
\end{eqnarray}
Here $t$, $U$, and $V$ denote the transfer integral 
between nearest-neighbor sites, 
the repulsive intra-site interaction, 
and the repulsive inter-site interaction between nearest-neighbor sites,
respectively. Furthermore,  
$a_i$ $(a_i^\dagger)$ is the annihilation (creation) operator
at the site $i$. 
In this study, we assume that the lattice is bipartite and consists of 
sublattices $A$ and $B$ and that the number of 
nearest-neighbor sites is $z$. 
The extended Bose--Hubbard model that  prohibits (allows) 
multiple boson occupations 
at one site is called the hard-core (soft-core)-extended Hubbard model.  


Several analytical and numerical methods exist
for studying the ground state of 
the extended Bose--Hubbard model. 
The strong-coupling perturbation theory \cite{Iskin} 
has been applied to obtain the phase boundary 
between the SF  and  non--SF phases.    
At least in the absence of the nearest-neighbor interaction 
\cite{Freericks1,Freericks2}, 
 the phase boundary determined by this theory agrees perfectly with that 
determined by quantum Monte Carlo (QMC) simulations 
\cite{Scalettar1,Niyaz1,Batrouni1,Otterlo1,Capogrosso1,Capogrosso2}.
However, this theory can neither distinguish between the SF and SS phases 
nor describe discontinuous transitions even if they exist. 
In contrast, the mean field (MF) approximation 
and equivalent Gutzwiller approximations 
\cite{Rokhsar,Krauth}  
can distinguish between the SF and SS phases and 
describe discontinuous transitions \cite{Kimura1,Kimura2,Krutitsky}, 
although their validity is limited for high dimensions. 

For the hard-core model at half filling ($N\ = \ 1/2$), both the MF approximation \cite{Matsuda,Liu,Scalettar2} for the mapped XXZ model 
and two-dimensional (2D) QMC studies \cite{Scalettar2,Batrouni2}
identified the first-order SF--solid transition and no SS phase. 
The MF approximation \cite{Matsuda} 
also showed that the transition from 
the SF phase to the checker-board solid phase is of the first order,  
and at the transition point $\mu\ = \ \mu^*$,  
the SS phase; solid phase; and PS phase in which
the SS phase and the solid phase coexistx are all degenerate. 

On the other hand, later QMC studies \cite{Batrouni3,Hebert1} 
showed that the SS phase is unstable against PS away from half filling. 

For the soft-core model, both 
Gutzwiller approximations \cite{Otterlo2,Scarola1}  
and QMC simulations \cite{Scalettar2,Batrouni2} 
have shown that checker-board SS phases can be stable against the PS. 
However, the details of the two types of study differ. 
The Gutzwiller studies showed stable SS phases 
for both broad filling and half filling, but 
the PS issue is not a concern in these cases. 
In contrast, the
2D QMC studies did not show the SS phase at half filling but identified 
SS phases above half filling; however, 
the system energy as a function of boson density 
was found to be concave below half filling, which indicates 
a possible PS instability \cite{Scalettar2}. 
A later 2D QMC study \cite{Sengupta1} 
showed more distinctly that the SS phase 
is unstable against PS below half filling 
by exhibiting the negative compressibility $\kappa\ = \ dN/d\mu<0$. 
This situation is somewhat similar in one dimension (1D).  
Both QMC \cite{Hebert2,Batrouni4} and density-matrix renormalization group 
\cite{Kuhner1,Kuhner2,Mishra1} studies have been performed in 1D.  
The SS phase was found only above half filling 
as in the 2D case but PS did not occur \cite{Batrouni4,Mishra1}. 

In a three-dimensional (3D) cubic lattice, 
Yamamoto and coworkers \cite{Yamamoto1} 
found some region of the parameter sets where the SS phase 
is stable below half filling in QMC calculations, 
which is consistent with  a MF phase diagram on the $N$--$t/U$ plane 
for $V\ = \ U/6$. 
Hence, it is interesting to study the 3D system further. 
On the other hand, other previous calculations 
assumed a grand canonical ensemble and 
did not directly include the possibility that 
the system can be separated into two phases. 
As a result, the phases comprising 
the separate phases were not directly shown. 
However, if we start with a canonical 
ensemble and explicitly include the possibility 
that the system can be separated into phases,  
we can precisely describe the phase diagram with a fixed 
boson number for the entire system. 

Such calculations might also be easily considered using the
grand canonical ensemble if we explicitly assume that 
the total free energy of the system 
is $E_{\rm tot}\ = \ \gamma E_{\rm SS}+(1-\gamma) E_{\rm sol}$.  
Here $\gamma$ is the ratio of the areas of the SS phases,  
and we neglect the surface energy between the SS and solid phases. 
However, if we minimize the free energy at the chemical potential $\mu$, 
we can obtain only $\gamma\ = \ 0$ or 1, but not 
an intermediate value $0< \gamma <1$ 
except for $\mu\ = \ \mu^*$ at which the SS and solid phases 
are degenerate [$E_{\rm SS}(\mu^*)\ = \ E_{\rm sol}(\mu^*)$]. 
That is, the solid and SS phases cannot be in the ground 
state at the same time,  
and a separate phase including two states cannot be obtained 
for $\mu\ \neq\ \mu^*$. 
To obtain a separate phase, 
we must tune the chemical potential from $\mu$ to $\mu^*$;  
furthermore, we must try to find a $\gamma$ 
that satisfies the boson number condition 
$\gamma N_{\rm SS}+(1-\gamma)N_{\rm sol}\ = \ N$.  
These calculations become more complicated when the number of possible phases 
is large (SF, SS, MI, and several solids). 
This issue, in which  PS can be obtained only 
on the phase boundary point(s) or curve(s) in the phase diagram 
but not for a finite region,  
generally occurs when we employ intensive 
variable(s) such as chemical potential 
in thermodynamics. 
The most popular case is the gas--liquid transition,  
where PS occurs only on the 
phase boundary curve in the pressure--volume plane for a given temperature. 
However, if we employ particle number, 
volume, and internal energy as thermodynamic variables, 
which are extensive quantities,  
we can obtain a finite region of PSs on the 
particle density--energy density plane. 


In this study, we study the ground-state properties of the 
extended Bose--Hubbard model on a bipartite lattice 
based on the Gutzwiller approximation 
in a canonical ensemble; hence, we do not have to 
tune the chemical potential of PS. 
We study a wide parameter regime and 
present three main figures for the phase diagram on the 
particle density--nearest-neighbor interaction plane.  
Our phase diagrams for small, intermediate, and large transfer integrals
differ significantly. 
The other main topic is the solid order parameter 
$\delta n\ = \ |N_A-N_B|$ [$N_{A(B)}$: the boson density on the 
$A(B)$ sublattice], which is also of interest 
and also depends strongly on the transfer integral.  
We employed the linear programming method \cite{numerical} 
to minimize the total energy for a particular boson number. 
Following this method, 
we can simultaneously determine the PS region, 
the phases that comprise the separate phase, 
and the ratio of each phase to the entire system. 

This paper is organized as follows, Section II 
introduces our calculation method, and  
Secs. III and IV present the hard-core and soft-core 
Bose--Hubbard models, respectively.   
Section V describes the effect of an 
improved calculation on the energy 
of the solid and MI phases. 
Section VI discusses the conclusions based on our results. 
Finally, the appendices explain the details of several perturbative 
calculations, the results of which are compared with the numerical 
results from the Gutzwiller approximation. 

\section{CALCULATION METHOD}

We employ the following Gutzwiller approximation
for the SS and SF phases: $\Psi\equiv \prod_i\Phi_i$. Here,
$\Phi_i$ is a variational wave function at the site $i$. 
We assume a bipartite lattice with sublattices $A$ and $B$ 
and a checker-board symmetry for the SS phase. 
The variational wave function 
is assumed to be $\Phi_i\ = \ \Phi_{A(B)}$ 
if the site $i$ belongs to the $A(B)$ sublattice. 
$\Phi_{A(B)}$ is written as a linear combination of states $|n\rangle$ 
with $n$ bosons as $\Phi_{A(B)}\ = \ \sum_{n}c_{A(B)n}|n\rangle$. 
The variational parameters $c_{A(B)n}$ are 
determined so as to minimize the energy expectation value. 
We use Powell's method \cite{numerical}  
to numerically optimize the variational parameters.  
If the result of the optimization shows 
$c_{An}\ \ne\ c_{Bn}$ and 
$c_{A(B)n}\ \ne\ \delta_{i(j)n}$, 
the phase is SS; however, if 
$c_{An}\ = \ c_{Bn}\ \ne\ \delta_{in}$, the phase is SF
($i$ and $j$ are non-negative and positive integers, 
respectively). 
The first inequality for the SS phase 
corresponds to the existence of a finite density difference 
between sublattices $A$ and $B$ $\delta n\ = \ |N_{A}-N_{B}|$
(i.e., the checker-board density order). 
Here $N_{A(B)}\ = \ \sum_n n|c_{A(B)n}|^2$ is the expectation value 
of the boson density at the $A(B)$ sublattice. 
The second inequality implies that 
the phase is not solid. For the soft-core model, 
in principle, the sum of $n$ must be taken from $n\ = \ 0$ 
to $\infty$; however, 
we take the sum from $n\ = \ 0$ to $n\ = \ 9$, 
which is sufficient for our calculation when $N\le 1$. 
For the hard-core model, by definition,  
we consider only the two states $|n\ = \ 0\rangle$ 
and $|n\ = \ 1\rangle$. 

The energy expectation value per site is calculated 
by the Gutzwiller variational wave function 
as 
\begin{eqnarray}
E&\ = \ &\langle H\rangle\nonumber\\
&\ = \ &\langle H_{\rm kin}\rangle+\langle H_{\rm int}^U\rangle
+\langle H_{\rm int}^V\rangle\nonumber\\
&\ = \ &-zt \prod_{i\ = \ A,B}\sum_n \sqrt{n+1} c_{in} c_{i(n+1)}\nonumber\\
&&+\frac{U}{4}
\sum_{i\ = \ A,B}\sum_n n(n-1)|c_{in}|^2
+\frac{zV}{2}N_A N_B. 
\end{eqnarray}
Furthermore, we assume that the energy per site of the MI phase 
with $N$ bosons per site is 
\begin{eqnarray}
E_{\rm MI}\ = \ \frac{U}{2}N(N-1)+\frac{zV}{2}N^2\ = \ \frac{zV}{2} 
\ \ \ {\rm for}\ N\ = \ 1\label{EMI}
\end{eqnarray}
and that of the solid phase with 
$N_A$ ($N_B$) bosons per site on the $A$($B$) sublattice is
\begin{eqnarray}
E_{\rm S}\ = \ \frac{U}{4}\sum_{i\ = \ A,B}N_i(N_i-1)+\frac{zV}{2}N_AN_B, \label{ES}
\end{eqnarray}
where $N_{A(B)}$ is a non-negative integer. 
In particular, for 
the solid phase when $N_A\ = \ 1$ and $N_B\ = \ 0$ (the S$_1$ phase),  
$E_{{\rm S}_1}\ = \ 0$; 
for the solid phase when $N_A\ = \ 2$ and $N_B\ = \ 0$ (the S$_2$ phase),  
$E_{{\rm S}_2}\ = \ U/2$. 
These energies correspond to those obtained from the Gutzwiller 
variational wave function with 
$c_{An}\ = \ c_{Bn}\ = \ \delta_{nN}$ for the MI phase
and $c_{An}\ = \ \delta_{nN_A}$ and $c_{Bn}\ = \ \delta_{nN_B}$ ($N_A\ \neq\ N_B$)
for the solid phase. 
Because we 
neglect the surface energy between different phases
by assuming the thermodynamic limit,  
the system's total energy per site is 
\begin{eqnarray}
E_{\rm tot}\ = \ \sum_{i} \gamma_i E_i
\end{eqnarray}
for the boson number condition
\begin{eqnarray}
N_{\rm tot}\ = \ \sum_{i} \gamma_i N_i. 
\end{eqnarray}
Here $\{E_i\}$ and $\{N_i\}$ (where $i\ = \ {\rm SF}$, SS, MI, and solids) 
are the energies and boson number densities,
respectively, of all possible phases and $\{\gamma_i\}$ 
represent the of area  
or volume ratio of the phase $i$ in the entire system. 
Following the linear programming method \cite{numerical}, we can 
minimize $E_{\rm tot}$ as a function of $\{\gamma_i\}$ \cite{note0}.  
Only one $\gamma_i$ value (corresponding to the 
uniform phase) or two $\gamma_i$ values (corresponding to PS) 
are automatically chosen to be nonzero as a result of minimizing $E_{\rm tot}$
because only one additional condition, the boson number condition, exists. 
For simplicity, however, we neglect the possibility 
of the separated phase that splits into 
the SF and SS phases, which is highly unlikely. Hereafter, we represent 
the separated phase consisting of phases X and Y as the PS(X + Y) phase.

\section{HARD-CORE MODEL} 

In the hard-core-extended Hubbard model, 
we prohibit multiple occupations at a site 
and define 
the Gutzwiller variational wave function as
\begin{eqnarray} 
\Phi_{A(B)}\ = \ c_{A(B)0}|0\rangle+c_{A(B)1}|1\rangle. \label{hardcore}
\end{eqnarray} 
At half filling, 
$|c_{A(B)0}|^2+|c_{A(B)1}|^2\ = \ 1$ 
according to the normalization condition of the wave function, 
and $|c_{A1}|^2+|c_{B1}|^2\ = \ 1$ according to the 
boson number condition. If we set $x\ = \ c_{A0}$, then 
$c_{A1}\ = \ c_{B0}\ = \ \sqrt{1-x^2}$ and $c_{B1}\ = \ x$. 
Hence, the system energy per site is 
\begin{eqnarray}
E&\ = \ &-zt c_{A0}c_{A1}c_{B0}c_{B1}
+\frac{zV}{2}|c_{A1}|^2|c_{B1}|^2 \nonumber\\
&\ = \ &\Big(zt-\frac{zV}{2}\Big)\Big[\Big(x^2-\frac{1}{2}\Big)^2-\frac{1}{4}\Big].
\end{eqnarray}
If $V>2t$, then $x^2\ = \ 0$ or $1$ and the phase is S$_1$;  
however, if $V<2t$, then $x^2\ = \ \frac{1}{2}$ and the 
phase is SF. Hence, the SF--S$_1$ phase transition occurs at $V\ = \ 2t$ 
and is discontinuous. 
This result agrees with those of previous studies \cite{Bruder1,Nozieres1}.

Away from half filling, 
we can obtain the SF--SS phase boundary by a perturbative calculation  
setting $c_{A(B)n}\ = \ c_n+\delta c_{A(B)n}$, 
where $c_n$ is the optimized value when 
we assume the SF phase ($c_{An}\ = \ c_{Bn}$) and 
$\delta c_{A(B)n}$ represents infinitesimal quantities that 
describe the possible SS instability (see Appendix A.1 for details). 
The result is 
\begin{eqnarray}
V_{\rm C}\ = \ t\ \frac{N^2+(1-N)^2}{N(1-N)}. \label{perturbation1}
\end{eqnarray}
That is, the energy of the SS phase is lower (higher) 
than that of the SF phase for $V>V_{\rm C}$ ($V<V_{\rm C}$). 
$V_{\rm C}^{\rm SS}$ is invariant under $N\leftrightarrow 1-N$ 
because of the particle--hole symmetry of the hard-core model. 
We also obtain the phase boundary between
the SF phase and the PS(SF + S$_1$) phase, which 
consists of the SF and the S$_1$ phases
by another perturbative calculation, assuming that the 
ratio of the S$_1$ phase to the entire system is infinitesimal 
(see Appendix A.2 for details). 
The result agrees with Eq. \ref{perturbation1}. 

To determine the ground state for $V>V_{\rm C}$, 
we numerically compared the energy of the SS and 
PS(SF + S$_1$) phases by using the Gutzwiller variational 
wave function as explained in the Introduction. 
We found that the PS(SF + S$_1$) and SS phases are 
degenerate for $V>V_{\rm C}$ within the numerical error.  
The phase diagram is shown in Fig. 1. 
The numerical phase boundary agrees with Eq. \ref{perturbation1}  
within the numerical error and the phase transitions  
are continuous as the perturbative calculation assumed.  
These results are consistent with those in Ref. \cite{Matsuda} which showed
that the SF--S$_1$ transition is of the first order 
and the SS, S$_1$, and PS(SF + S$_1$) phases are all degenerate 
at the transition point $\mu\ = \ \mu^*$.  

We also confirmed that both $\gamma_{\rm S_1}$ 
(the ratio of the S$_1$ phase) in the PS(SF + S$_1$) phase  
and $\delta n\ = \ |N_A-N_B|$ in the SS phase 
are finite for $V>V_{\rm C}$ [Figs. 2(a) and 2(b)].  
As $N$ approaches $1/2$, 
$\gamma_{\rm S_1}$ increases rapidly as a function of $(V-V_{\rm C})/V_{\rm C}$,
because at half filling, no SS phase occurs and 
the SF--S$_1$ phase transition is discontinuous. 

\begin{figure}
\includegraphics[height= 2.5in]{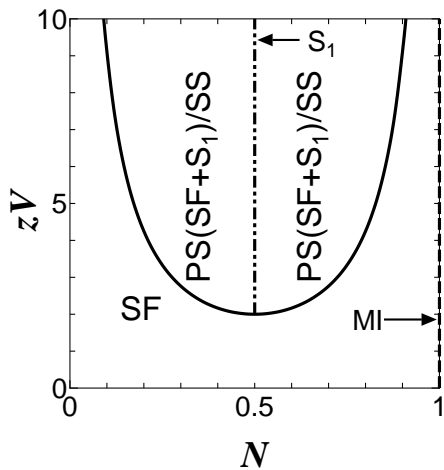}
\caption{
Phase diagram of the hard-core model. 
In the PS(SF + S$_1$)/SS phase, the SS and 
PS(SF + S$_1$) phases are degenerate.  
The two-dot-dashed line at half filling indicates the solid phase; 
the dashed line at unit filling indicates the MI phase. 
}
\label{fig1}
\end{figure}

\begin{figure}
  \subfigure[]{
    \includegraphics[height=2.5in]{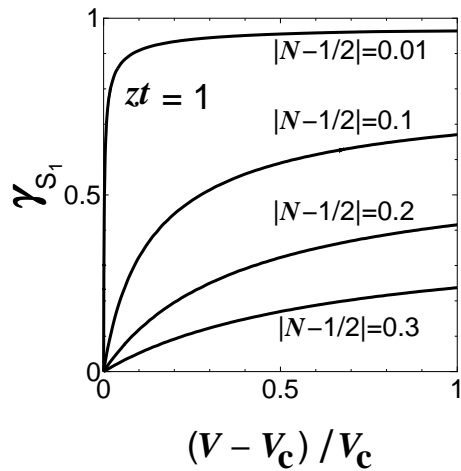}
  \label{fig2a}}
  \subfigure[]{
    \includegraphics[height=2.5in]{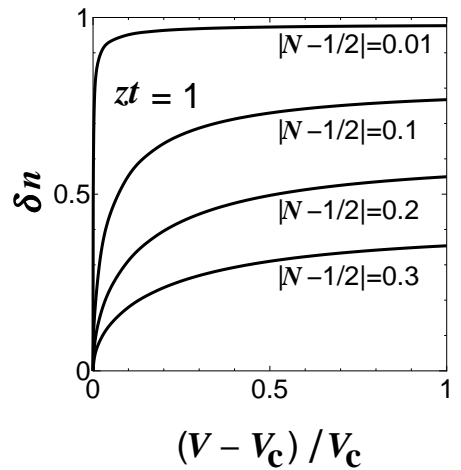}
  \label{fig2b}}
\caption{
(a) $V$ dependence of 
$\gamma_{\rm S_1}$ [the ratio of the S$_1$ phase 
in the PS(SF + S$_1$) phase] for $zt\ = \ 1$. 
(b) $\delta n\ = \ |N_A-N_B|$ 
in the SS phase for $zt\ = \ 1$. 
$V_{\rm C}$ is the critical value for the 
SF--PS(SF + S$_1$) or SF--SS transition at each value of $N$ 
($zV_{\rm C}\ = \ 4.250$, 2.762, 2.167, 2.002 for 
$|N-1/2|\ = \ 0.3$, 0.2, 0.1, 0.01, respectively).   
}
  \label{fig2}
\end{figure}

\section{SOFT-CORE MODEL}

In this section, we study the soft-core-extended Hubbard model.
We employ the Gutzwiller variational wave function and 
optimize its variational parameters numerically. 
Hereafter, we call this the full numerical calculation(s);   
when we do not refer to the calculation method, 
the result was obtained by the full numerical calculations. 
We also perform perturbative calculations that 
limit the Hilbert space of the Gutzwiller variational 
wave function and include partial numerical calculations. 
We compare these perturbative calculations with 
the full numerical calculations. 
In subsection A, we examine 
the SS phase between the SF 
and solid phases at half filling. 
In subsection B, we examine the case away from half filling, 
which is the main part of this paper. 
Here we show that both the phase diagram on the $zV/U-N$ plane 
and $\delta n\ = \ |N_A-N_B|$ change 
qualitatively from small $zt/U$ to large $zt/U$. 

\subsection{HALF FILLING}
In this subsection, we primarily examine the SS phase at half filling.  
We compare the critical value of $V$
for the SS--S$_1$ transition with that of the SF--SS transition
because if the former is larger than the latter, we can 
obtain the SS phase between the SF and S$_1$ phases. 

By a perturbative calculation with 
an infinitesimal SS component added to the S$_1$ phase, 
we obtain the critical value of $V$ for the SS--S$_1$ transition 
by a power-law expansion of $zt/U$:
\begin{eqnarray}
zV_{\rm C}^{\rm SS-S_1}\ = \ 2zt+2\frac{z^2t^2}{U}+2\frac{z^3t^3}{U^2}+O\Big(\frac{z^4t^4}{U^3}\Big) \label{eq3}
\end{eqnarray}
(see Appendix B.1 for details). 
We can also obtain the critical 
value of $V$ for the SF--SS transition 
\begin{eqnarray}
zV_{\rm C}^{\rm SF-SS}\ = \ 2zt+2\frac{z^2t^2}{U}+O\Big(\frac{z^4t^4}{U^3}\Big),
\label{eq4}
\end{eqnarray}
by another perturbative calculation, in  
which an infinitesimal $\delta n\ = \ |N_A-N_B|$ is added to the SF phase
(see Appendix B.2 for details). 
Because $V_{\rm C}^{\rm SF-SS}<V_{\rm C}^{\rm SS-S_1}$,  
the SS phase is possible for 
intermediate $V$ satisfying $V_{\rm C}^{\rm SF-SS}<V<V_{\rm C}^{\rm SS-S_1}$. 

We verified these results numerically. 
Figure 3 shows the phase boundaries obtained 
by the above-mentioned perturbative calculations 
(dashed and dot-dashed curves) and 
the full numerical calculations (solid curves). 
A stable SS phase appears between the SF 
and S$_1$ phases. Both the SS--S$_1$ and SF--SS phase boundaries 
determined by perturbative calculations 
agree well with those determined by the full numerical calculations
(especially at small $zt/U$ as expected). 
Figure 4 shows the interaction dependence of $\delta n\ = \ |N_A-N_B|$.  
Here, $zV_{\rm C}\ = \ zV_{\rm C}^{\rm SF-SS}$ is the critical value 
for the SF--SS transition at each value of $zt/U$. 
A finite $\delta n$ less than unity 
shows the density order of the SS phase, 
whereas $\delta n\ = \ 0(1)$ corresponds to the SF(S$_1$) phase. 
We found that $\delta n$  continuously becomes finite at 
the SF--SS phase transition. It 
changes more rapidly for smaller $zt/U$. 
This demonstrates that 
the SS phase disappears and the SF-S$_1$ discontinuous transition occurs
in the hard core limit $zt/U\rightarrow 0$. 

On the other hand, 2D QMC calculations at half filling \cite{Scalettar2,Batrouni2} 
showed that the SF phase directly transitions to the S$_1$ phase 
and no SS phase appears. 
However, $zt/U$ might not be sufficiently large to allow the SS phase 
to be easily found there, and a QMC simulation (as a matter of course, 
not only 2D but also 3D) for a large $zt/U$ 
has a possibility for finding the SS phase. 

\begin{figure}
\includegraphics[height= 2.5in]{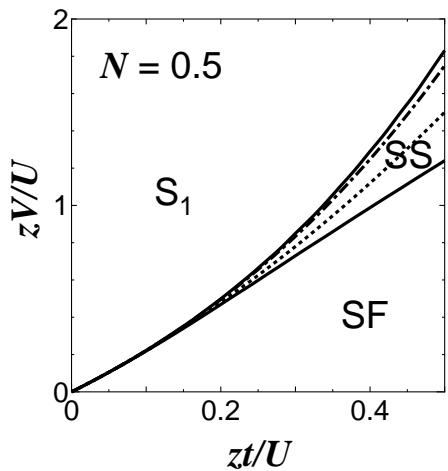}
\caption{
Phase diagram of the soft-core model at half filling. 
Solid curves represent numerical SF--SS and SS--S$_1$ phase boundaries.  
Dashed (dot-dashed) curve represents the SF--SS 
(SS--S$_1$) phase boundary obtained 
by a perturbative calculation using Eq. \ref{eq4} (Eq. \ref{eq3}). 
}
\label{fig3}
\end{figure}

\begin{figure}
\includegraphics[height= 2.5in]{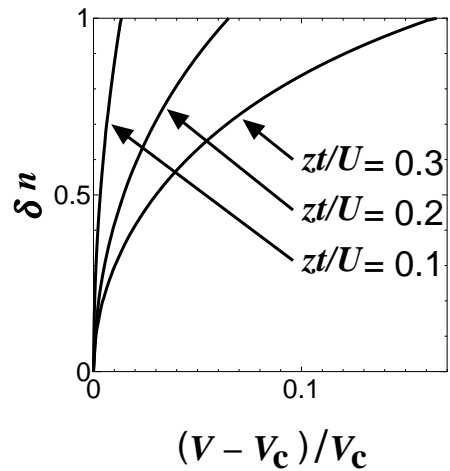}
\caption{
$V$ dependence of the density difference 
between sublattices $A$ and $B$ $\delta n\ = \ |N_A-N_B|$ in the SS phase
of the soft-core model at half filling. 
$V_c$ is the critical value for the 
SF--SS transition at each value of $zt/U$ 
($zV_{\rm C}/U\ = \ 0.219$, 0.469, 0.730 for 
$zt/U\ = \ 0.1$, 0.2, 0.3, respectively).   
}
\label{fig4}
\end{figure}

\subsection{AWAY FROM HALF FILLING}
In this section, 
for comparison with the results of the full numerical calculations, 
we calculate the results of the following three perturbative 
equations for the phase boundary (see Appendices A, B.2, and B.3 for details). 
Perturbation 1 is Eq. \ref{perturbation1}, which 
assumes a limited Hilbert space with $|N\ = \ 0\rangle$ and $N\ = \ 1\rangle$ 
and is the same as that used in Sec. III; 
perturbation 2 is Eq. \ref{perturbation2}, which yields
the SF--SS phase boundary; and 
perturbation 3 is Eq. \ref{perturbation3}, which yields
the SF--PS(SF + S$_1$) or SF--PS(SF + S$_2$) phase boundary. 
For perturbations 2 and 3, 
we employ a limited Hilbert space with $|N\ = \ i\rangle$, where 
$i\ = \ 0$, 1, and 2. 
Hence, results obtained by 
perturbations 2 and 3 are expected to be better than those obtained 
by perturbation 1. 
However, perturbations 2 and 3 require 
a numerical calculation to 
minimize the SF energy in the limited Hilbert space. 

We begin by analyzing the phase diagram for a 
small transfer integral $zt/U$. 
Figure 5(a) shows the phase diagram for $zt/U\ = \ 0.03$,   
which resembles  that of the hard-core model. 
Namely, the upper bound of $zV/U$ for the SF phase 
has a minimum value near half filling, 
which is similar to that of the hard-core model
with particle-hole symmetry. 
PS occurs for large $zV/U$ 
below half filling, and 
the separated phase is the PS(SF + S$_1$) phase, as in the hard-core model. 
However, the SS phase appears above half filling. 
These properties agree with those obtained in 
a 2D QMC study with $zt/U\ = \ 0.08$ \cite{Sengupta1}. 

Figure 5(a) plots the result of perturbation 1 (dot-dashed curve). 
It agrees almost perfectly with the full numerical calculation
(solid curve) except at large $N$ where the SF--SS phase 
 boundary disappears because the PS(SS + S$_2$) phase appears there 
[see also Fig. 5(b), which expands Fig. 5(a) around $zV/U\ = \ 1$].  
Both perturbation 2 for the SF--SS boundary curve and 
perturbation 3 for the SF--PS(SF + S$_1$) boundary curve 
yield almost the same results as perturbation 1. 
Therefore, they also agree almost exactly 
with the full numerical calculations except at large $N$. 

Figure 5(b) shows that PSs also appear above half filling 
but the region is very small compared to that obtained by 
2D QMC \cite{Sengupta1}. 
The appearance of the PS(SS + S$_2$) phase and the intricate structure of the 
phase diagram are non-trivial;  
however, the two SS phases that  sandwich 
the PS(SS + S$_2$) phase have different characteristics: 
one having a smaller zV/U resembles the SF phase and 
the other having a larger zV/U resembles the S$_2$ phase, 
as we will see below 
through $\delta n\ = \ |N_{\rm A}-N_{\rm B}|$. 
Namely,  SF, SS, S$_2$, and the PS comprising 
these phases are almost 
degenerate for $zV/U\sim 1$ above half filling.

\begin{figure}
  \subfigure[]{
    \includegraphics[height=2.5in]{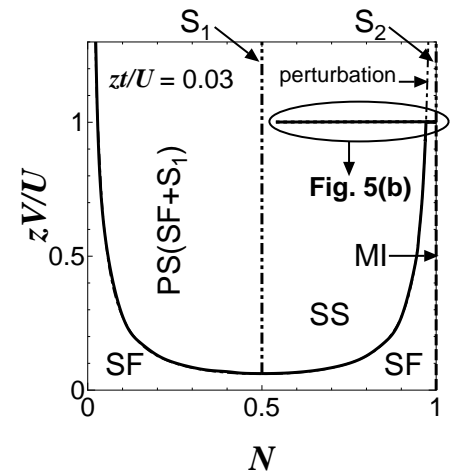}
  \label{fig5a}}
  \subfigure[]{
    \includegraphics[height=2.5in]{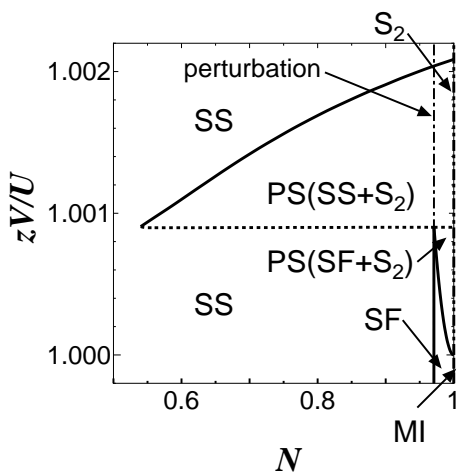}
  \label{fig5b}}
  \caption{
Phase diagram of the soft-core model for $zt/U\ = \ 0.03$. 
(a) Entire phase diagram. 
Solid curve represents the SF--PS(SF + S$_1$) or SF--SS phase boundary. 
Dot-dashed curve for the SF--SS phase boundary 
obtained by the perturbative calculation in Eq. \ref{perturbation1} 
cannot be distinguished from the solid curve  
except at large $N$, where the SF--SS transition curve disappears. 
(b) Expansion of (a) around $zV/U\ = \ 1$ and $N\ = \ 1$. 
Solid (dashed) curves for the phase boundaries 
show continuous (discontinuous) phase transitions. 
In both (a) and (b), the two-dot-dashed lines 
at half filling and unit filling 
indicate the solid phases (S$_1$ and S$_2$, respectively), 
and the long-dashed line at unit filling 
indicates the MI phase.
} 
  \label{fig5}
\end{figure}

Figure 6 confirms that the SS phase overcomes the PS(SF + S$_1$) phase 
above half filling because the critical value of the nearest-neighbor 
interaction $V_{\rm C}$ for the SF--SS transition 
is always smaller than that for the virtual SF--PS(SF + S$_1$) transition. 
Here, the virtual SF--PS(SF + S$_1$) transition was obtained 
by setting the variational parameter of the 
Gutzwiller variational wave function to $c_{An}\ = \ c_{Bn}$. 
The difference between the curves for the SF--SS transition and 
the virtual SF--PS(SF + S$_1$) transition 
disappears in the hard-core limit of $zt/U\rightarrow 0$,    
where the SS and PS(SF + S$_1$) phases are degenerate,  
as discussed in Sec. III. 

\begin{figure}
  \includegraphics[height=2.5in]{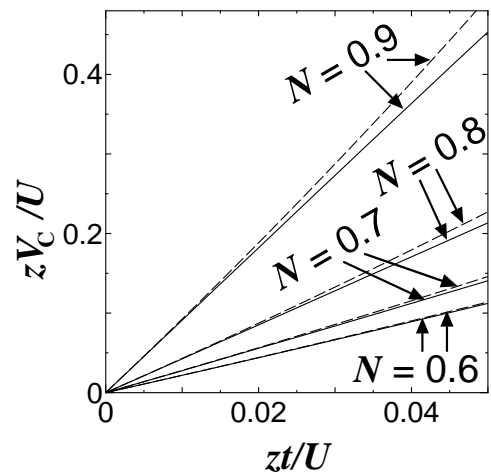}
  \caption{
Critical nearest-neighbor interaction $V_{\rm C}$ 
for the SF--SS transition (solid curves)
and the virtual SF--PS(SF + S$_1$) transition (dashed curves)
above half filling. 
Note that both solid and dashed curves appear 
for $N\ = \ 0.6$, although they are very close. 
}
\label{fig6}
\end{figure}

Figure 7(a) shows the $V$ dependence of $\delta n$ obtained 
by the full numerical calculation. 
Figure 7(b) shows an expansion of the region around $1.001\le zV/U\le 1.002$. 
Note that $\delta n$ has a discontinuity at $zV_{\rm C}/U\simeq 1.001$, 
because the discontinuous SS--PS(SS + S$_2$) transition occurs there. 
Interestingly, the SS phase exhibits a small $\delta n$ for $zV/U\leq 1$, 
whereas it exhibits a large $\delta n\simeq 2N$ ($N_A\simeq 2N$ and 
$N_B\simeq 0$) for $zV/U\ge 1.002$ 
after a rapid increase in $\delta n$ in a narrow PS(SS + S$_2$) region
($1.001\leq zV/U\leq 1.002$), where the curves of $\delta n$ 
for different $N$ are almost indistinguishable 
[see Fig. 7(b)]. 
Because  $\delta n\simeq 2N$, 
$\delta n$ is larger for larger $N$ at $zV/U\ge 1.002$. 
In contrast, $\delta n$ is smaller for larger $N$ for $zV/U\le 1$, 
demonstrating that the critical value of $zV/U$ for the SF--SS transition 
is an increasing function of $N$ above half filling. 
The fact that $\delta n\simeq 2N$ for $zV/U\gtrsim 1$ indicates that 
the SS phase is similar to  the S$_2$ phase, in which 
$N_{\rm A}\ = \ 2$ and $N_{\rm B}\ = \ 0$. The fact that 
$\delta n\simeq 0$ for $zV/U\lesssim 1$ indicates that 
the SS phase resembles  the SF phase. 
Note that $zV/U\ = \ 1$ is also the phase transition point 
for $t\ = \ 0$ at unit filling: the MI phase (i.e., no-density-order phase) 
for $zV/U<1$ becomes 
the S$_2$ phase (density order phase) for $zV/U>1$. That is, 
even for a finite but small $t$ 
and $N<1$, the characteristics of the ground state 
change rapidly near $zV/U\simeq 1$. 

\begin{figure}
  \subfigure[]{
    \includegraphics[height=2.5in]{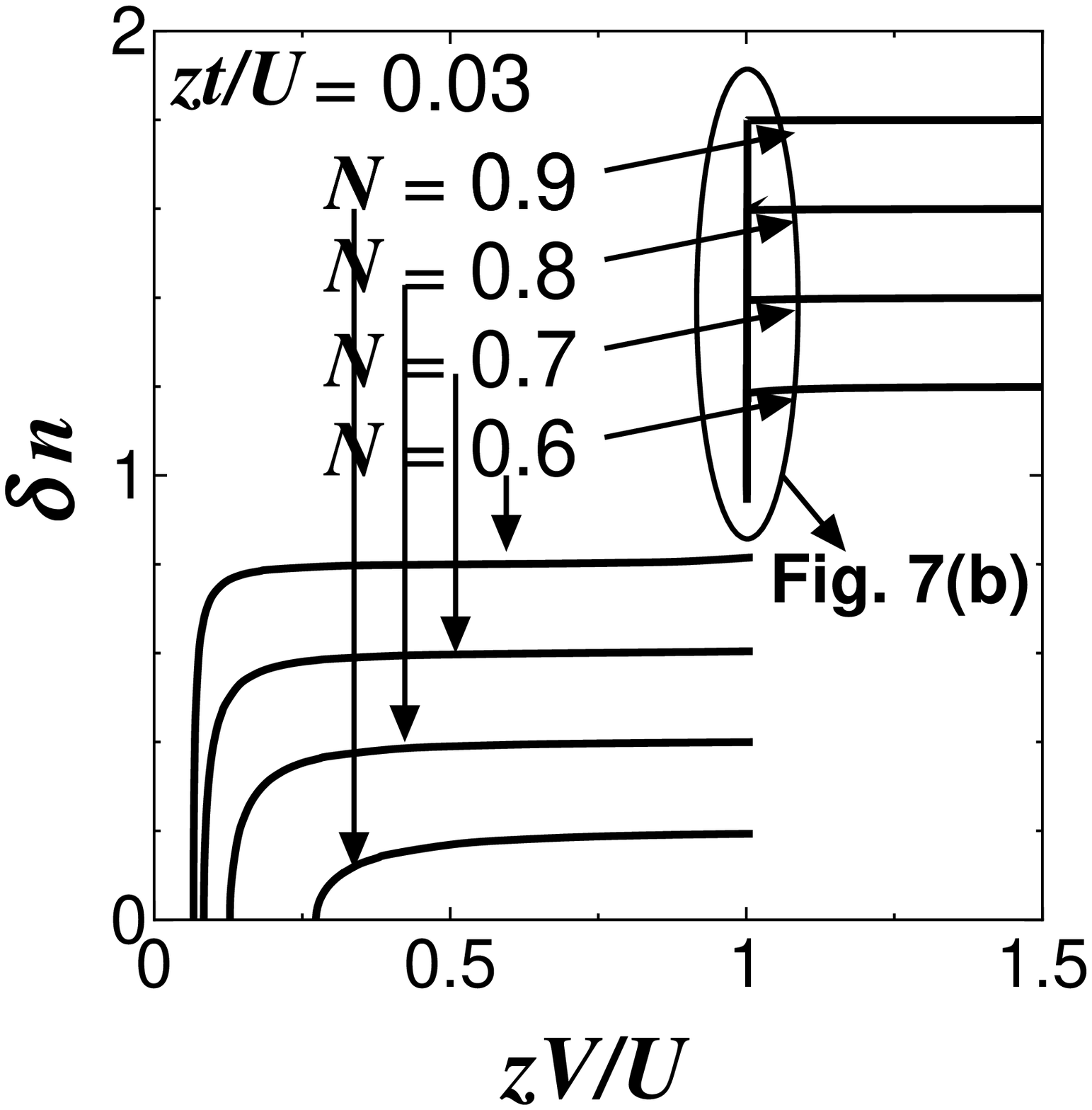}
  \label{fig7a}}
  \subfigure[]{
    \includegraphics[height=2.5in]{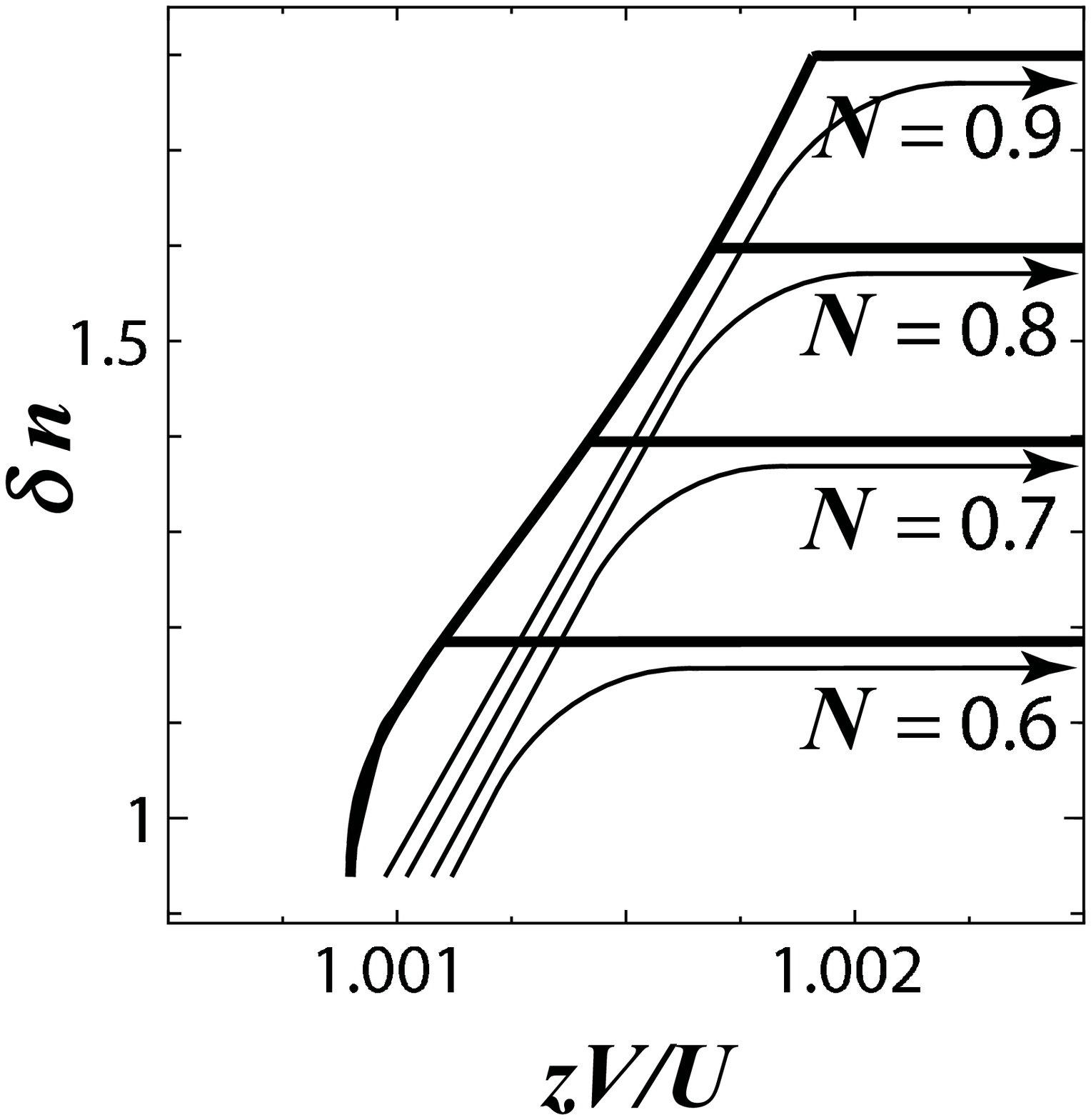}
  \label{fig7b}}
  \caption{
$V$ dependence of 
$\delta n\ = \ |N_A-N_B|$ in the SS phase
of the soft-core model for $zt/U\ = \ 0.03$. 
$\delta n$ is larger at smaller $N$ for $zt/U
\le 1.000$,  
 discontinuous at $zV_{\rm C}/U\simeq 1.001$, 
and smaller at smaller $N$ for $zt/U\ge 1.002$.    
(b) Expansion around the PS(SS + S) phase 
($1.001\leq zV_{\rm C}/U\leq 1.002$). 
In the PS(SS + S) phase, $\delta n$ is a sharply  
increasing function of $zV/U$, and the curves of $\delta n$ 
for different $N$ cannot be distinguished. 
Thus, the curved arrows describe the increase in $\delta n$ at 
each value of $N$ when we enlarge $zV/U$. 
}
\end{figure}

Next, we study the case of an intermediate $zt/U$ value of 0.3.  
Figures 8(a) and 8(b) show the phase diagram;  
the PS region above half filling 
exists, but again is very small. 
The phase diagram clearly departs from that 
of the hard-core model with the particle-hole symmetry. 
Interestingly, for an intermediate $zV/U$, the SS phase 
overcomes the PS even below half filling. This agrees with 
a recent 3D QMC simulation with $zt/U\ = \ 0.33$ \cite{Yamamoto1}.  
Figure 9 shows $\delta n$ ($\gamma_{\rm S_1}$) 
below half filling in the SS  (PS(SF + S$_1$)) phase. 
The SS--PS(SF + S$_1$) transition is discontinuous:  
for $N\ = \ 0.3$, $0.35$, and $0.4$, $\delta n$ ($\gamma_{\rm S_1}$)
is finite (zero) 
in the SS phase and discontinuously becomes zero (finite) 
in the PS(SF + S$_1$) phase. 
For $N\ = \ 0.25$, $\delta n$ is zero in the entire figure 
because the SS phase does not exist. 

Returning to Fig. 8(a), 
we see that the results from 
perturbative calculations (dot-dashed curves) 
agree well with those from the full numerical 
calculations (solid curves).   
Assuming the S$_1$ phase for the solid, perturbation 3 
is in almost perfect agreement 
with the SF--PS(SF + S$_1$) phase boundary below half filling.  
Perturbation 2 is also in excellent agreement with 
the SF--SS phase boundary except for 
the region of large $N\sim 1$ where the SF--SS transition 
becomes discontinuous, and thus  
cannot be described by the perturbative calculation. 

Figure 8(b) shows an enlargement of Fig. 8(a) around $zV/U\ = \ 1.1$ and $N\sim 1$, 
where a small PS(SF + S$_2$) phase appears. 
The curve of perturbation 3 in Fig. 8(b), which is calculated 
assuming the S$_2$ phase for the solid, 
seems to be  far from the results obtained by 
the full numerical calculations; however, 
the difference is in fact approximately 1$\%$ at most.

\begin{figure}
  \subfigure[]{
    \includegraphics[height=2.5in]{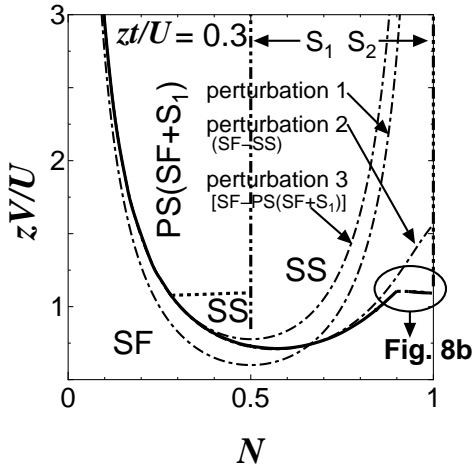}
  \label{fig8a}}
  \subfigure[]{
    \includegraphics[height=2.5in]{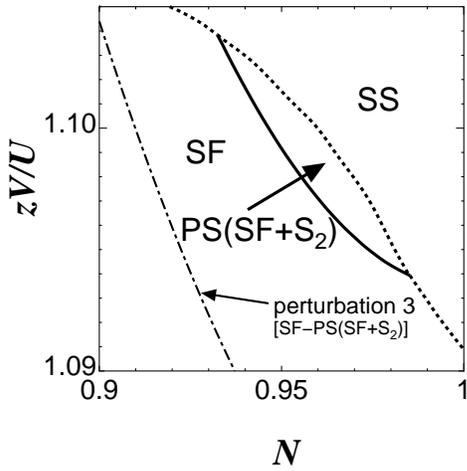}
  \label{fig8b}}
  \caption{
Phase diagrams of the soft-core model for $zt/U\ = \ 0.3$. 
(a) shows Entire phase diagram. 
Solid curve shows the continuous SF--PS(SF + S$_1$) and SF--SS transitions. 
Dot-dashed curves 
are the results of the perturbative calculations  
Eq. \ref{perturbation1} (perturbation 1), 
Eq. \ref{perturbation2} (perturbation 2), and Eq. \ref{perturbation3} (perturbation 3) for the SF--PS(SF + S$_1$) transition. 
Two-dot-dashed lines at half filling and
unit filling represent the solid phases (S$_1$ and S$_2$, respectively). 
(b) Expansion of the region  around $zV/U\ = \ 1$ and $N\ = \ 1$. 
Solid curve shows the continuous 
SF--PS(SF + S$_2$) phase transition; dashed curve
shows the discontinuous SF--SS and PS(SF + S$_2$)--SS phase transitions. 
Dot-dashed curve shows Eq. \ref{perturbation3} (perturbation 3)
for the SF--PS(SF + S$_2$) transition.  
} 
  \label{fig8}
\end{figure}

\begin{figure}
  \includegraphics[height=2.5in]{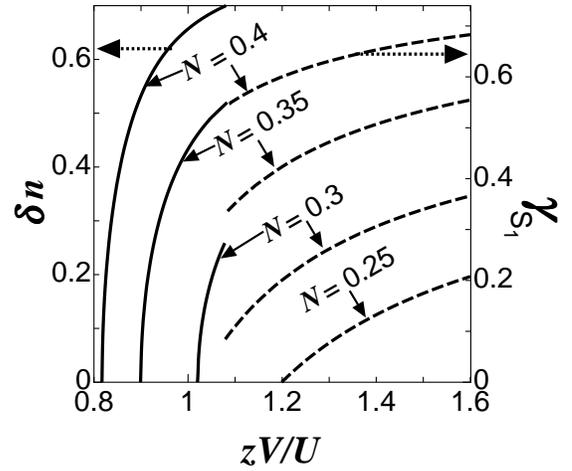}
  \caption{
$V$ dependences of 
$\delta n\ = \ |N_A-N_B|$ (solid curves) and $\gamma_{\rm S_1}$  (dashed curves)
in the soft-core model for $zt/U\ = \ 0.3$ below half-filling.  
}
\label{fig9}
\end{figure}

Figure 10 shows the $V$ dependence of 
$\delta n$ above half filling for $zt/U\ = \ 0.3$. 
As in Fig. 7 for $zt/U\ = \ 0.03$, 
$\delta n\simeq 2N$ is larger for larger $N$ at $zV/U>1.1$, whereas
$\delta n$ is smaller for larger $N$ for $zV/U<1.1$. 
Hence, as in the case of small $zt/U$, 
the characteristics of the SS change drastically from SF-like 
to solid (S$_2$)-like around $zV/U\sim 1$ when we increase $zV/U$. 
Note that in contrast to the case of $zt/U\ = \ 0.03$, 
$\delta n$ shows no discontinuity except at $N\ = \ 0.9$ 
because the SF--SS transition is continuous. 
As a result, the two $\delta n$ curves for two different $N$ values 
intersect smoothly around $zV/U\simeq 1.1$ except at $N\ = \ 0.9$.  
For $N\ = \ 0.9$, $\delta n$ is zero for  $zt/U<1.105$ 
(because the phase is SF, at which $\delta n\ = \ 0$) and 
discontinuously becomes finite for $zt/U>1.105$,  
because the SF--SS transition is discontinuous there.  

\begin{figure}
  \includegraphics[height=2.5in]{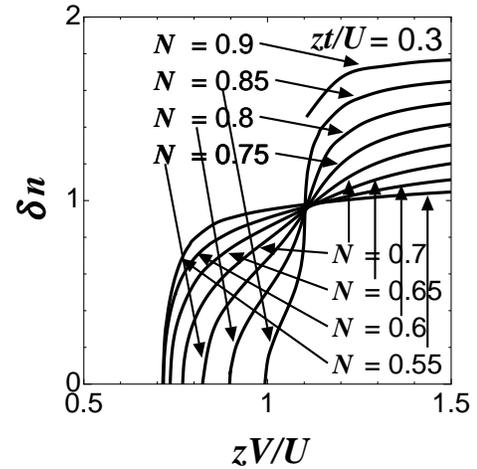}
  \caption{
$V$ dependence of 
$\delta n\ = \ |N_A-N_B|$ of the soft-core model for $zt/U\ = \ 0.3$ above half filling. 
}
\label{fig10}
\end{figure}

Finally, we examine  a large $zt/U$($\ = \ 1$). 
Figure 11 shows the phase diagram. 
Only the SF and SS phases exist, and there are no PSs. 
The SF--SS transition is continuous throughout the figure. 
The critical $zV/U$ value between these phases is a decreasing function 
of the boson density $N$. This may be explained as follows.   
For large $zt/U$, the ratio $t$ to $V$ is the only important factor  
determining the phase (SF or SS, not the PS), 
because the lattice does not play an important role except 
at half or unit filling, and $N$ affects only 
the SF--SS phase boundary ($V$ 
affects the phase more strongly for larger $N$). 
As a result, no rapid change in the ground state properties 
occurs around $zV/U\ = \ 1$ in contrast to the case of small or intermediate $zt/U$.   

The dot-dashed curve of perturbation 1 
does not agree with the full numerical calculation (solid curve) 
because perturbation 1 exhibits
the particle-hole symmetry of the hard-core model, which  
is distinctly broken here. However, the 
dot-dashed curve of perturbation 2 is in excellent agreement 
with the solid curve except at large $N$($\sim 1$).  

\begin{figure}
    \includegraphics[height=2.5in]{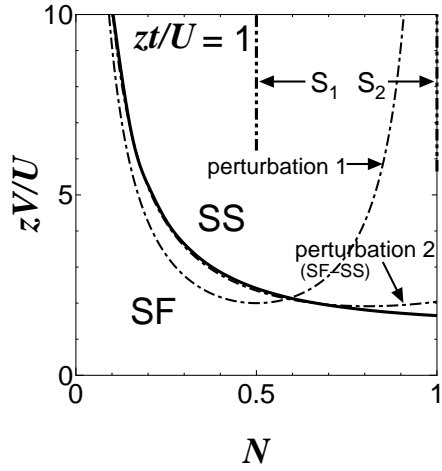}
  \caption{
Phase diagram of the soft-core model for $zt/U\ = \ 1$. 
Solid curve represents the SF--SS phase boundary at which 
the phase transition is continuous. 
Dot-dashed curves 
show the results of perturbation 1 (Eq. \ref{perturbation1}) 
and 2 (Eq. \ref{perturbation2}). 
Two-dot-dashed lines at half filling and 
unit filling show the S$_1$ and S$_2$ phases, respectively. 
} 
  \label{fig11}
\end{figure}

Figure 12 shows the $V$ dependence of 
$\delta n$ in the SS phase of the soft-core model for $zt/U\ = \ 1$. 
$\delta n$ is larger for larger $N$ at the same $zV/U$ 
because the critical value of $zV/U$ for the SF--SS transition 
is smaller for larger $N$ (Fig. 11). 
Unlike the cases of $zt/U\ = \ 0.03$ (Fig. 6) or $zt/U\ = \ 0.3$ (Fig. 10), 
$\delta n$ is a smooth increasing function 
of $N$ and $zV/U$, and the 
two curves of $\delta n$ for two different $N$ values
no longer intersect. Note that 
$\delta n$ does not change rapidly 
around $zV/U\sim 1$ as expected from the phase diagram (Fig. 11).

\begin{figure}
  \includegraphics[height=2.5in]{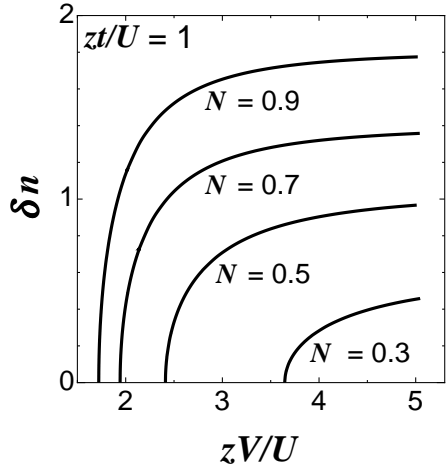}
  \caption{
$V$ dependence of 
$\delta n\ = \ |N_A-N_B|$ in the soft-core model at $N\ = \ 0.3$, 0.5, 0,7, and 0.9 for $zt/U\ = \ 1$. 
}
\label{fig12}
\end{figure}

\section{Effect of improved calculation 
on the energy of the solid and MI phases}
The energy of the solid and MI phases employed in the previous sections 
may be significantly higher than the exact energy. 
Hence, in this section, 
we improve the calculated energies in these phases
by employing the perturbation theory \cite{Iskin} up to the order of  
$t^2/U$ or $t^2/V$ 
as 
\begin{eqnarray}
E_{\rm MI}&\ = \ &\frac{zV}{2}-\frac{2zt^2}{U-V},\\
E_{\rm S_1}&\ = \ &-\frac{zt^2}{2(z-1)V},\\
E_{\rm S_2}&\ = \ &\frac{U}{2}-\frac{zt^2}{(2z-1)V-U}. 
\end{eqnarray}
We can obtain the phase diagram by employing these 
improved energies. 
Hereafter, we assume $z\ = \ 6$ for the 3D cubic lattice. 
However, note that this improvement may be excessively favorable 
for the solid and MI phases and unfavorable for the SF and SS phases. 
Therefore, a correction  on the order of $t^2/U$ or $t^2/V$ should also be 
applied to the energies of the SF and SS phases 
close to the SF (or SS)--solid (or MI) phase boundary. 
In addition, the denominators of the equations describing 
$E_{\rm MI}$, $E_{{\rm S}_1}$, and $E_{{\rm S}_2}$ 
diverge for $V\ = \ U$, $V\ = \ 0$ and $(2z-1)V\ = \ U$, respectively. 
Hence, we exclude the S$_1$ phase around $V\ = \ 0$ and 
the S$_2$ phase around $zV/U\ = \ z/(2z-1)\ = \ 6/11$. 
We also neglect the PS into two solids (S$_1$ and S$_2$) 
which is unlikely but indeed appears in this approximation. 
Figures 13 and 14 show the phase diagrams 
for $zt/U\ = \ 0.03$ and 
$zt/U\ = \ 0.3$, respectively. 
(The phase diagram for $zt/U\ = \ 1$ is the same as that in Fig. 11.)
In both figures, the regions of separated phases become large,  
as expected. For $zt/U\ = \ 0.03$, the PS 
into the SF and MI phases [PS(SF + MI) phase] appears. Furthermore,  
for $zt/U\ = \ 0.3$, the SS phase disappears below half filling 
because of existence of the PS(SF + S$_1$) phase. 
These results suggest that the solid and MI energies 
have somewhat over-improved. 
On the other hand, the regions of  PS into 
the SS and solid phases become large above half filling, 
which resembles the results of a 2D QMC study by Sengupta and coworkers \cite{Sengupta1}.
\begin{figure} 
    \includegraphics[height=2.5in]{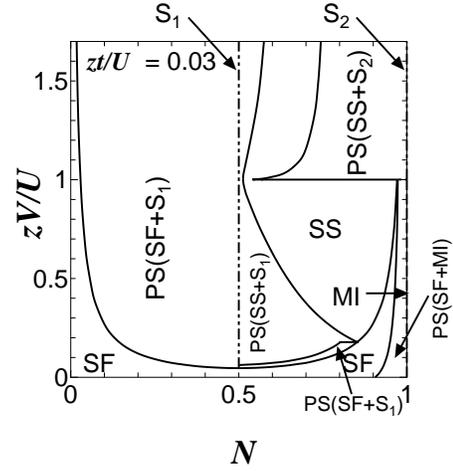}
  \caption{
Phase diagram of the soft-core model 
for $zt/U\ = \ 0.03$. 
Energies of the solid and MI phases are improved (see text). 
} 
  \label{fig13}
\end{figure}
\begin{figure}
    \includegraphics[height=2.5in]{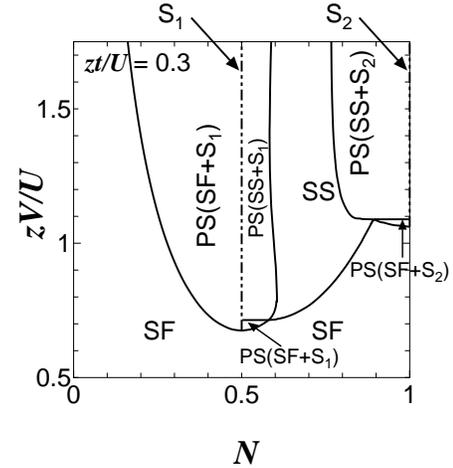}
  \caption{
Phase diagram of the soft-core model 
for $zt/U\ = \ 0.3$. Energies of the solid 
and MI phases are improved (see text). 
} 
  \label{fig14}
\end{figure}

\section{Conclusion}
In this work, we studied the hard-core and soft-core-extended 
Hubbard models by the Gutzwiller variational 
wave function. We adopted a canonical ensemble and
a linear programming method to include PSs more directly
in our calculations. 
In the hard-core model away from half filling, 
we showed that the PS(SF + S$_1$) and SS phases are degenerate 
above a critical value of the nearest-neighbor interaction $V$, 
which is consistent with a previous MF study \cite{Matsuda}. 

Unlike the hard-core model,  
the soft-core model at half filling has  
a possible SS phase  between the solid and SF phases
and all the phase transitions are continuous.  

Away from half filling, the phase diagram 
depends drastically on the transfer integral $t$. 
For small $zt/U$, the shape of the SF 
region is similar to that of the hard-core model. 
The PS(SF + S$_1$) phase appears below half filling 
and the SS phase appears only above half filling,  
as in the 2D QMC studies.
For intermediate $zt/U$, the SS phase appears close to the 
PS below half filling  
as in the 3D QMC study. 
The phase diagram becomes  simpler for large $zt/U$, where 
only the continuous SF--SS phase transition appears, 
and the critical value of $zV/U$ at the phase boundary  
is a smooth decreasing function of $N$. 

The nearest-neighbor interaction 
dependence of $\delta n\ = \ |N_A-N_B|$, 
which shows the density wave order of the SS phase,
is also interesting. 
For a small $zt/U$ of 0.03, $\delta n$ is a discontinuous 
function of $zV/U$; furthermore, 
the SS phase has a small $\delta n$ for small $zV/U$($\lesssim 1$) 
and large $\delta n$($\simeq 2N$)  
($N_A\simeq 2N$ and $N_B\simeq 0$) for large $zV/U$($\gtrsim 1$). 
In addition, 
$\delta n$ is larger for smaller (larger) $N$ 
for $zV/U\lesssim 1$ ($zV/U\gtrsim 1$). 
For an intermediate $zt/U$ of 0.3, 
the behavior of $\delta n$ is similar to that for a small n
$zt/U\ = \ 0.03$. 
In detail, however, unlike the case of $zt/U\ = \ 0.03$,  
the two curves of $\delta n$ for two different $N$ values continuously
intersect around $zV/U\simeq 1.1$ because $\delta n$ 
is a continuous increasing function of $zV/U$ except at large $N$. 
For a large $zt/U\ = \ 1$, $\delta n$ is a smooth increasing 
function of $N$ and $zV/U$, and the 
two curves of $\delta n$ for two different $N$ values  
no longer intersect.  

Throughout this paper, 
we found that our  perturbative calculations 
determined the phase boundary curves very well   
except for large $N$($\sim  1$). 

We also studied the effects of the improved perturbative calculation
 on the energy of the solid and MI phases. The improvement enlarges the 
region of PS into the SS and solid phases above half filling, 
which resembles  the results of the 2D QMC study. However, the improvement is 
excessively favorable for the solid and MI phases, 
resulting in an unusual enlargement of the PSs. Therefore,
the energies of the SF and SS phases should also be improved in future work. 

Because the Gutzwiller approximation is not precise,  
exact numerical calculations such as QMC simulations are needed
to check our results. 
Although some details (such as the very complicated 
phase diagrams for $N\sim 1$ and small $zt$) 
might be artifacts of our approximation, 
we believe that the important results in 
the phase diagrams and $\delta n$ values in the SS phase 
are  worth studying in detail. 
For instance, the transfer integral dependence of the 
entire phase diagram seems to remain an open question 
not only in the 3D case but also in the 2D case, 
especially for the large $zt/U$ regime, in which  
 an SS phase below or at half filling might exist.

\begin{acknowledgments}
I sincerely thank D. Yamamoto and I. Danshita
for fruitful discussions. 
\end{acknowledgments}

\appendix

\section{PERTURBATIVE CALCULATIONS FOR THE HARD-CORE MODEL}

\subsection{SF--SS TRANSITION}
In the Gutzwiller approximation, 
the energy expectation value
of the Hamiltonian (Eq. \ref{Hamiltonian})
is obtained as 
\begin{eqnarray} 
E&\ = \ &E_{\rm kin}+E_{\rm int}^{V},\nonumber\\
E_{\rm kin}&\ = \ &-ztc_{A0}c_{B0}c_{A1}c_{B1},\nonumber\\
E_{\rm int}^V&\ = \ &\frac{zV}{2}c_{A1}^2c_{B1}^2. 
\end{eqnarray}
Here, we assumed that $c_{A(B)n}$ $(n\ = \ 0,1)$ is real 
without the loss of generality and used 
$\langle a_{A(B)}\rangle\ = \ c_{A(B)0}c_{A(B)1}$ 
and $\langle a_{A(B)}^\dagger a_{A(B)}\rangle\ = \ c_{A(B)1}^2$. 
To consider a possible SS phase infinitesimally 
close to the SF--SS phase boundary, 
we set $c_{A(B)n}\ = \ c_n+\delta\alpha_n(\delta\beta_n)$, 
where $\delta\alpha_n$ and $\delta\beta_n$ are infinitesimal quantities. 
The normalization condition of the wave function 
for sublattices $A$ and $B$ is written as 
\begin{eqnarray}
(c_0+\delta \alpha_0)^2+(c_1+\delta\alpha_1)^2&\ = \ &1,\nonumber\\
(c_0+\delta \beta_0)^2+(c_1+\delta\beta_1)^2&\ = \ &1.
\end{eqnarray}
The boson number condition is written as
\begin{eqnarray}
(c_1+\delta\alpha_1)^2+(c_1+\delta\beta_1)^2\ = \ 2N. 
\end{eqnarray}
These equations can be rewritten  using the 
normalization ($\sum_nc_n^2\ = \ 1$) and boson number 
conditions ($\sum_n n c_n^2\ = \ N$)  in the SF phase 
as
\begin{eqnarray}
4c_0\delta x_0+\delta x_0^2 +\delta y_0^2&\ = \ &0,\nonumber\\
4c_1\delta x_1+\delta x_1^2 +\delta y_1^2&\ = \ &0,\nonumber\\
c_0\delta y_0+c_1\delta y_1+\delta x_0 \delta y_0&\ = \ &0.\label{d} 
\end{eqnarray}
Here we introduced new variables 
$\delta x_n(\delta y_n)\ = \ \delta\alpha_n+(-)\delta\beta_n$ and 
$\delta y_n\ \ne \ 0$. 
These equations can be further rewritten in the lowest order as
\begin{eqnarray}
4c_0\delta x_0+\delta y_0^2&\ = \ &0,\nonumber\\
4c_1\delta x_1+\delta y_1^2&\ = \ &0,\nonumber\\
2c_0\delta y_0+2c_1\delta y_1&\ = \ &0.
\end{eqnarray}
Hence, the value of $\delta x_n$ is of the same order as $\delta y_n^2$, and 
$\delta x_n$ can be rewritten in terms of $\delta y_n^2$. 
The energy expectation values can also be rewritten as
\begin{eqnarray} 
E_{\rm kin}&\ = \ &-zt\Big[N(1-N)-\frac{1}{2}(1-N)\delta y_1^2 
-\frac{1}{2}N\delta y_0^2\Big],\nonumber\\ 
E_{\rm int}^V&\ = \ &\frac{zV}{2}\Big[N^2-N\delta y_1^2\Big]. 
\end{eqnarray} 
By using Eq. \ref{d}, 
\begin{eqnarray} 
E&\ = \ &-zt\ N(1-N)+\frac{zt}{2}N^2\nonumber\\ 
&&+\frac{1}{2}\Big[zt\frac{N^2+(1-N)^2}{1-N}-zVN\Big]\delta y_1^2. 
\end{eqnarray} 
If the coefficient of $\delta y_1^2$ is positive (negative),  
the phase is the SF (SS). 
By setting  the coefficient of $\delta y_1^2$ 
equal to zero, we obtain the critical value of $V$ 
for the SF--SS transition 
\begin{eqnarray}
V_{\rm C}\ = \ t\ \frac{N^2+(1-N)^2}{N(1-N)}.
\end{eqnarray}
This is nothing but Eq. \ref{perturbation1}. 

\subsection{PHASE SEPARATION INTO SF AND SOLID OR MOTT PHASE}

To study the PS from the SF phase into the SF phase and 
the S$_1$ (MI) phase, we set the ratio
of the S$_1$ (MI) phase as $\gamma_{\rm SM}$. 
When the total boson density is $N$, 
the number density condition 
is written as
\begin{eqnarray}
(1-\gamma_{\rm SM})(N-\delta N)+\gamma_{\rm SM} N_{\rm SM}\ = \ N. \label{eqa}
\end{eqnarray}
Here $N_{\rm SM}\ = \ N_{\rm S_1}\ = \ 1/2$ ($N_{\rm SM}\ = \ N_{\rm MI}\ = \ 1$). 
If $\delta N\ = \ \gamma_{\rm SM}\ = \ 0$ the phase is the uniform SF phase. 
In contrast, if $\delta N,\gamma_{\rm SM}\ \ne\ 0$, the phase is the PS 
consisting of the SF and S$_1$ (MI) phases, and
$N-\delta N$ is the number density of the SF phase. 
From Eq. \ref{eqa}, we obtain
\begin{eqnarray}
\delta N\ \simeq\ (N_{\rm SM}-N)\gamma_{\rm SM}  \label{eqb}
\end{eqnarray}
near the SF--PS phase boundary ($\delta N\ll N$ and $\gamma_{\rm SM}\ll 1$).  
Because $|c_1|^2\ = \ N-\delta N$ and $|c_0|^2\ = \ 1-|c_1|^2$, 
the system energy per site is 
\begin{eqnarray} 
E&\ = \ &\gamma_{\rm SM} E_{\rm SM}+(1-\gamma_{\rm SM})E_{\rm SF}, \nonumber\\
E_{\rm SF}&\ = \ &-zt|c_0|^2|c_1|^2+\frac{zV}{2}|c_1|^4\nonumber\\
&\ = \ &-zt(1-N+\delta N)(N-\delta N)\nonumber\\
&&+\frac{zV}{2}(N-\delta N)^2. \label{eqc}
\end{eqnarray}
Here, $E_{\rm SF}$ is the energy of the SF phase and
$E_{\rm SM}$ is that of the S$_1$ (MI) phase: 
$E_{\rm SM}\ = \ E_{\rm S_1}\ = \ 0$ ($E_{\rm SM}\ = \ E_{\rm MI}\ = \ zV/2$).  
To obtain the phase boundary, we substitute Eq. \ref{eqb} into Eq. \ref{eqc}. 
The result is 
\begin{eqnarray}
E&\ = \ &-zt\ N(1-N)+\frac{zV}{2}N^2\nonumber\\
&&+\Big\{ztN^2+\frac{zV}{2}N^2+E_{\rm SM}\nonumber\\
&&+\big[-zVN+zt(1-2N)\big]N_{\rm SM}\Big\}\gamma_{\rm SM} 
\end{eqnarray} 
in the lowest order of $\gamma_{\rm SM}$. 
For the PS that splits into 
the SF and S$_1$ phases, $N_{\rm SM}\ = \ 1/2$,
and $E_{\rm SM}\ = \ 0$. 
If the coefficient of $\gamma_{\rm SM}$ is positive (negative),  
the phase is the SF (PS). 
By setting the coefficient of $\gamma_{\rm SM}$
to zero, we obtain the critical value of $V$ 
for the PS. The result is the same as Eq. \ref{perturbation1}. 
In contrast, for the PS that splits into 
the SF and  MI phases, $N_{\rm SM}\ = \ 1$, $E_{\rm SM}\ = \ zV/2$,  
and the coefficient of $\gamma_{\rm SM}$ is positive definite:
\begin{eqnarray}
\Big(zt+\frac{zV}{2}\Big)(N-1)^2>0.
\end{eqnarray} 
Hence, the PS into the SF and MI phases 
does not occur. 

\section{PERTURBATIVE CALCULATIONS FOR THE SOFT-CORE MODEL}
\subsection{SS--S$_1$ TRANSITION AT  HALF FILLING}
To obtain the SS--S$_1$ phase boundary at half filling (Eq. \ref{eq3}), 
we set $c_{A0}\ = \ \delta \alpha_0$, $c_{A1}\ = \ 1-\delta \alpha_1$, 
$c_{A2}\ = \ \delta\alpha_2$, $c_{B0}\ = \ 1-\delta\beta_0$, 
$c_{B1}\ = \ \delta\beta_1$, and $c_{B2}\ = \ \delta\beta_2$
infinitesimally close to the phase boundary. 
If $\delta\alpha_n\ = \ \delta\beta_n\ = \ 0$ $(n\ = \ 0,1,2)$, the phase is SF,  
but if $\delta\alpha_n,\delta\beta_n\ \ne\  0$, the phase is SS. 
The normalization conditions of the wave function for sublattices $A$ and $B$ 
lead to the following equations in the lowest order: 
\begin{eqnarray}
\delta\alpha_1&\ = \ &\frac{1}{2}(\delta \alpha_0^2+\delta \alpha_2^2),\nonumber\\
\delta\beta_0&\ = \ &\frac{1}{2}(\delta \beta_1^2+\delta \beta_2^2). 
\end{eqnarray}
In addition, the boson number condition leads to 
\begin{eqnarray}
\delta \alpha_1\ = \ \delta\alpha_2^2
+\frac{1}{2}\delta \beta_1^2+\delta\beta_2^2. 
\end{eqnarray}
From these equations, we can eliminate $\delta\alpha_0$, 
$\delta\alpha_1$, and $\delta\beta_0$ and obtain the 
energy in terms of $\delta\alpha_2$, 
$\delta\beta_1$, and $\delta\beta_2$ as
\begin{eqnarray}
E&\ = \ &\frac{U}{2}\delta\alpha_2^2+\frac{zV}{2}\delta\beta_1^2
+\Big(\frac{U}{2}+zV\Big)\delta\beta_2^2\nonumber\\
&&-zt\Big(\sqrt{2}\delta\alpha_2
+\sqrt{\delta\alpha_2^2+\delta\beta_1^2+2\delta\beta_2^2}\Big)
\delta\beta_1\nonumber\\
&\ = \ &\Big[\frac{U}{2}x^2+\frac{zV}{2}+\Big(\frac{U}{2}+zV\Big)y^2\nonumber\\
&&-zt\Big(\sqrt{2}x+\sqrt{1+x^2+2y^2}\Big)\Big]\delta\beta_1^2, \label{z}
\end{eqnarray}
where we have set $\delta\alpha_2\ = \ x\delta\beta_1$ 
and $\delta\beta_2\ = \ y\delta\beta_1$.
The minimization conditions of Eq. \ref{z} are written as 
\begin{eqnarray}
&&\frac{1}{\delta\beta_1^2}\frac{\partial E}{\partial x}
\ = \ Ux-\sqrt{2}zt-\frac{ztx}{\sqrt{1+x^2+2y^2}}\ = \ 0,\ \ \ \ \\
&&\frac{1}{\delta\beta_1^2}\frac{\partial E}{\partial y}
\ = \ (U+2zV)y-\frac{2zty}{\sqrt{1+x^2+2y^2}}\ = \ 0.\ \ \ \ 
\end{eqnarray}
The latter equation leads to $y\ = \ 0$,  
because we assume small $zt$($\ll U$). From the former equation, we have
\begin{eqnarray}
x\ = \ \frac{\sqrt{2}zt}{U}\Big(1+\frac{zt}{U}\Big)+O\Big(\frac{z^3t^3}{U^3}\Big). 
\end{eqnarray}
By substituting the $x$ value obtained above into Eq. \ref{z}, we 
obtain the optimized coefficient of $\delta \beta_1^2$. 
If it is positive (negative),  
the phase is S$_1$ (SS). By setting it to zero, we obtain 
the critical value of $zV$ for the SS--S$_1$ phase transition (Eq. \ref{eq3}).

\subsection{SF--SS TRANSITION}
As explained in the text, we limit the Hilbert space 
to three states $|0\rangle$, $|1\rangle$, and
$|2\rangle$. Furthermore, we set $c_{An}\ = \ c_n+\delta\alpha_n$ and 
$c_{Bn}\ = \ c_n+\delta\beta_n$
where $\delta\alpha_n$ and $\delta\beta_n$ 
($n\ = \ 0$, 1, 2) are infinitesimal quantities. 
We determine the $c_n$ ($n\ = \ 0$, 1, 2) value required 
to minimize the energy of the SF phase ($\delta\alpha_n\ = \ \delta\beta_n\ = \ 0$). 
As in the  hard-core model (Appendix A.1), 
we introduce 
$\delta x_n(\delta y_n)\ = \ \delta\alpha_n+(-)\delta\beta_n$, 
where $\delta y_n\ \ne\ 0$ for the SS phase. 
The wave function normalization 
and boson number conditions lead to
\begin{eqnarray}
&&c_0\delta x_0+c_1\delta x_1+c_2\delta x_2
\ = \ -\frac{1}{4}(\delta y_0^2+\delta y_1^2+\delta y_2^2),\ \ \ \ \nonumber\\
&&2c_1\delta x_1+4c_2\delta x_2\ = \ -\frac{\delta y_1^2}{2}
-\delta y_2^2,\ \ \ \ \nonumber\\
&&c_0\delta y_0+c_1\delta y_1+c_2\delta y_2\ = \ 0\ \ \ \ 
\end{eqnarray}
in the lowest order of $\delta x_n$ and $\delta y_n$
($\delta x_n$ are of the order of $\delta y_n^2$). 
By eliminating $\delta x_0 $, $\delta x_1$, and $\delta y_1^2$ in  
the above equations, we can write 
the energy per site in terms of $\delta x_2$, 
$\delta y_0$, and $\delta y_2$ as  
\begin{eqnarray}
E&\ = \ &E_{\rm kin}+E_{\rm int}^U+E_{\rm int}^{V},\nonumber\\
E_{\rm kin}&\ = \ &-zt\Big[c_1^2\big(\sqrt{2} c_2+c_0\big)^2
+X \delta x_2\nonumber\\
&&-\frac{1}{4}\big(a_0\delta y_0^2
+a_{02}\delta y_0\delta y_2+
a_2\delta y_2^2\big) \Big],\nonumber\\
E_{\rm int}^U&\ = \ &U\Big[c_2^2+c_2\delta x_2+
\frac{1}{4}\delta y_2^2\Big],\nonumber\\
E_{\rm int}^V&\ = \ &\frac{zV}{2}\Big[\big(c_1^2+2c_2^2\big)^2
-\big(c_0\delta y_0-c_2\delta y_2\big)^2\Big], \ \ \ \label{x}
\end{eqnarray}
where 
\begin{eqnarray}
X&\ = \ &\big(\sqrt{2} c_2+c_0\big)
\Big[
c_1^2\Big(\sqrt{2}+\frac{c_2}{c_0}\Big)\nonumber\\
&&-2c_2\big(\sqrt{2}c_2+c_0\big)
\Big],\\
a_0&\ = \ &\Big(2+\frac{\sqrt{2}c_2}{c_0}\Big)c_1^2
+\frac{2c_0^2}{c_1^2}
\big(\sqrt{2} c_2+c_0\big)^2,\\
a_{02}&\ = \ &2\Big[
\sqrt{2}c_1^2+\frac{2c_0c_2}{c_1^2}
\big(\sqrt{2} c_2+c_0\big)^2\Big],\\
a_2&\ = \ &2\big(\sqrt{2} c_2+c_0\big)^2
\Big(1+\frac{c_2^2}{c_1^2}\Big)\nonumber\\ 
&&+c_1^2\Big(1-\frac{\sqrt{2}c_2}{c_0}\Big).
\end{eqnarray}
The total coefficient of $\delta x_2$ for $E$ in Eq. \ref{x} 
is zero, because the variation in $\delta x_n$ corresponds to 
that in $c_n$ in the SF phase 
and the values of $c_n$ were already determined to minimize $E$.
To be concrete, the energy of the SF phase is 
represented by $c_2$ when $\delta y_n\ = \ 0$ as 
\begin{eqnarray}
E_{\rm SF}&\ = \ &-zt(N-2c_2^2)\Big(\sqrt{2}c_2+\sqrt{1-N+c_2^2}\Big)^2\nonumber\\
&&+\frac{zV}{2}N^2+Uc_2^2,\label{e1}
\end{eqnarray}
where $c_0$ and $c_1$ were already eliminated 
by the the wave function normalization
and boson number conditions. 
The minimization condition $dE_{\rm SF}/dc_2\ = \ 0$
leads to 
\begin{eqnarray}
Uc_2 &\ = \ & zt\Big[ c_2 \left(\sqrt{2}c_2+\sqrt{1-N+c_2^2}\right)^2 \nonumber\\
&&+\frac{N-2c_2^2}{\sqrt{1-N+c_2^2}}
\left( \sqrt{2}c_2+\sqrt{1-N+c_2^2} \right) \nonumber\\
&& \times \left(c_2+\sqrt{2(1-N+c_2^2)}\right) \Big], \label{a}
\end{eqnarray}
and we can easily verify that the total coefficient 
of $\delta x_2$ for $E$ in Eq. \ref{x} is 
zero. $E$ is further rewritten as
\begin{eqnarray}
E&\ = \ &{\rm const.}+A\delta y_0^2+B\delta y_0 \delta y_2
+C\delta y_2^2, \nonumber\\
A&\ = \ &\frac{zt}{4}a_0-\frac{zV}{2}c_0^2,\nonumber\\
B&\ = \ &\frac{zt}{4}a_{02}+zVc_0c_2,\nonumber\\
C&\ = \ &\frac{zt}{4}a_{2}-\frac{zV}{2}c_2^2+\frac{U}{4}.
\end{eqnarray}
Hence, $\delta y_n$ becomes finite and the phase becomes SS 
when $B^2>4AC$, which corresponds to  
$V>V_{\rm C}^{\rm SF-SS}$, where
\begin{eqnarray}
V_{\rm C}^{\rm SF-SS}&\ = \ &\frac{t}{8}
\frac{zt(4a_0a_2-a_{02}^2)+4Ua_0}
{zt(c_2^2a_0+c_0c_2a_{02}+c_0^2a_2)+Uc_0^2}. \ \ \label{perturbation2}
\end{eqnarray} 
To obtain the numerical value of Eq. \ref{perturbation2}, we 
numerically calculate the value of $c_2$ that minimizes $E$ (Eq. \ref{e1}) 
and determine $c_0$ by using the normalization 
and boson number conditions. By 
replacing $c_0$ and $c_2$ in Eq. \ref{perturbation2},  
we obtain the critical value $V_{\rm C}^{\rm SF-SS}$. 

On the other hand, at half filling, 
we analytically obtain a power-law expansion of $zt/U$ for 
$V_{\rm C}^{\rm SF-SS}$ (Eq. \ref{eq4}).  
We start from the 
minimization condition obtained from Eq. \ref{e1} 
up to the second order of $c_2$:  
\begin{eqnarray}
\frac{dE}{dc_2}
\ = \ -zt(1+c_2-9c_2^2)+2Uc_2\ = \ 0.  
\end{eqnarray}
From the above equation, we obtain $c_2\ = \ zt/(2U)+z^2t^2/(4U^2)$. 
We determine $c_0$ and $c_1$ by $c_2$ through the 
normalization and boson number conditions. 
Now, $a_0$, $a_{02}$, and $a_2$ are written as 
\begin{eqnarray}
a_0&\ = \ &2+5\delta +13\delta^2,\nonumber\\
a_{02}&\ = \ &\sqrt{2}(1+2\delta+6\delta^2),\nonumber\\
a_2&\ = \ &\frac{3}{2}(1+2\delta+6\delta^2), 
\end{eqnarray}
respectively, where $\delta\ = \ zt/(2U)$. By substituting 
$a_0$, $a_{02}$, and $a_2$ into Eq. \ref{perturbation2}, 
we obtain Eq. \ref{eq4}.  

\subsection{PHASE SEPARATION INTO THE SF AND SOLID OR MOTT PHASE}

As explained in the text and in Appendix B. 2, 
we limit the Hilbert space 
to three states: $|0\rangle$, $|1\rangle$, and 
$|2\rangle$. We assume that $c_n\ = \ c_n^{(0)}$ (where $n\ = \ 0$, 1, 2) is 
determined to minimize the total energy of the SF phase 
(not the separated phase). 
The wave function normalization and boson number conditions 
are written as 
$\sum_nc_n^{(0)2}\ = \ 1$ and $\sum_nnc_n^{(0)2}\ = \ N$, respectively. 
For the separated phase, we set 
$c_i\rightarrow c_n^{(0)}+\delta c_n$ and 
$N\rightarrow N-\delta N$ for the SF phase 
in the separated phase 
and $\gamma_{\rm SM}$ as the 
ratio of the solid (or MI) phase to the entire system. 
As in a similar calculation in Appendix A.2, 
the boson number condition in the entire system
is written as $\delta N\ = \ \gamma_{\rm SM}(N_{\rm SM}-N)$ in the lowest 
order of $\gamma_{\rm SM}$, 
where $N_{\rm SM}\ = \ 1/2$ (1) 
for the solid S$_1$ (S$_2$) phase  
and $N_{\rm SM}\ = \  1$ for the MI phase. 
The normalization and boson number conditions  
for the SF phase in the separated phase are 
written as $\sum_n(c_n^{(0)}+\delta c_n)^2\ = \ 1$ and 
$\sum_nn(c_n^{(0)}+\delta c_n)^2\ = \ N-\delta N$, respectively. 
These equations are rewritten as
\begin{eqnarray}
\delta c_0&\ = \ &-\frac{1}{4c_0}\big(2c_1\delta c_1-\delta N\big),
\nonumber\\
\delta c_2&\ = \ &-\frac{1}{4c_2}\big(2c_1\delta c_1+\delta N\big). 
\end{eqnarray}
From these equations and the minimization condition for $E_{\rm SF}$ 
(Eq. \ref{a}), we determine the energy of the SF phase $E_{\rm SF}$ 
in the separated phase as 
\begin{eqnarray}
E_{\rm SF}&\ = \ &-zt(N-2c_2^2)\Big(\sqrt{2}c_2+\sqrt{1-N+c_2^2}\Big)^2\nonumber\\
&&+\frac{zV}{2}N^2+Uc_2^2+Y\delta N, 
\end{eqnarray}
where
\begin{eqnarray}
Y&\ = \ &-\frac{zt}{2}c_1^2(\sqrt{2}c_2+c_0)\Big(-\frac{\sqrt{2}}{c_2}+\frac{1}{c_0}\Big)\nonumber\\
&&-zVN-\frac{U}{2}.
\end{eqnarray}
Note that the $\delta c_1$ term is eliminated:
the optimization process of $E_{\rm SF}$ is similar to that in Appendix B.2,  
and the coefficient of $\delta c_1$ is found to be zero by  Eq. \ref{a}. 
Because the ratio of the SF phase in the separated phase 
is $1-\gamma_{\rm SM}$ and $\delta N\ = \ (N_{\rm SM}-N)\gamma_{\rm SM}$, 
the energy of the entire system $E$ is written as 
\begin{eqnarray}
E&\ = \ &E_{\rm SF}(1-\gamma_{\rm SM})+E_{\rm SM}\gamma_{\rm SM} \nonumber\\
&\ = \ &-zt(N-2c_2^2)\Big(\sqrt{2}c_2+\sqrt{1-N+c_2^2}\Big)^2 \nonumber\\
&&+\frac{zV}{2}N^2+Uc_2^ 2\nonumber\\
&&+(Azt-BzV-CU+E_{\rm SM})\gamma_{\rm SM},
\end{eqnarray}
where 
\begin{eqnarray}
A&\ = \ &c_1^2\Big[(\sqrt{2}c_2+c_0)^2
+\frac{1}{2}\Big(1+\sqrt{2}\frac{c_0^2-c_2^2}{c_0c_2}\Big)
(N_{\rm SM}-N)\Big],\nonumber\\
B&\ = \ &N\Big(N_{\rm SM}-\frac{N}{2}\Big), \nonumber\\
C&\ = \ &c_2^2+\frac{1}{2}(N_{\rm SM}-N),  
\end{eqnarray}
and $E_{\rm SM}$ is the energy of the solid (MI) phase:
$E_{\rm SM}\ = \ 0$ ($E_{\rm SM}\ = \ U/2$) for the S$_1$ (S$_2$) phase, 
and $E_{\rm SM}\ = \ zV$ for the MI phase. 
If the coefficient of $\gamma_{\rm SM}$ is positive (negative), 
the phase is SF (PS). 
By setting the coefficient of $\gamma_{\rm SM}$ to zero, we find that 
the critical value of $V$ for the SF--PS transition is 
\begin{eqnarray}
V_{\rm C}^{\rm SF-PS}\ = \ \frac{ztA-UC+E_{\rm SM}}{zB}. \label{perturbation3}
\end{eqnarray}


\begin{thebibliography}{18}
\bibitem{note0}
In more detail, to obtain $E_{\rm SF(SS)}$
for a given $N_{\rm SF(SS)}$, 
we add a term to the Hamiltonian 
$\beta(N-N_{\rm SF(SS)})^2$, 
where $\beta$ is a large positive value
and is an additional variational parameter 
for minimizing the total energy. 
By minimizing the expectation value of $H+\beta(N-N_{\rm SF(SS)})^2$, 
we obtain a SF phase with a given $N_{\rm SF(SS)}$ 
without the direct boson number condition 
$\sum_{n} n|c_{in}|^2\ = \ N_{\rm SF}$ 
($\sum_{n}\sum_{i\ = \ A,B} n|c_{in}|^2\ = \ 2N_{\rm SS}$):       
Although we tried to directly impose the boson number condition 
$\sum_{n} n|c(n)|^2\ = \ N_{\rm SF}$ on the wave function 
in addition to the normalization condition $\sum_{n} |c(n)|^2\ = \ 1$, 
the convergence of the calculation was poor. 
Because $N_{\rm new}\ = \ \sum_{n} n|c(n)|^2$  
is slightly different from the original $N_{\rm SF(SS)}$ 
as a result of the minimization,  
we adopt $N_{\rm new}$ as the boson density. 
By changing $N_{\rm SF(SS)}$, which gives $N_{\rm new}$ and 
$E_{\rm SF(SS)}\ = \ \langle H\rangle$, 
we finally obtain the minimized $E_{\rm tot}$ and the corresponding 
phases and their ratios  $\{\gamma_i\}$.
\end{thebibliography}
\end{document}